\documentclass[onecolumn,superscriptaddress,secnumarabic,amssymb,amsmath,nobibnotes,aps,prd,nofootinbib,altaffilletter]{revtex4}
\usepackage{amsmath,amssymb}
\usepackage{graphicx}
\usepackage{dcolumn}
\usepackage{bm,color}
\usepackage{hyperref}
\usepackage{accents}
\usepackage{amssymb,float}
\usepackage{amsmath}
\usepackage{multirow}
\usepackage{siunitx}
\usepackage{tabularx}
\usepackage{booktabs}
\usepackage{url}
\usepackage{orcidlink}
\usepackage{soul}
\usepackage{mathrsfs}

\usepackage{mathtools}

\DeclarePairedDelimiterX\braket[2]{\langle}{\rangle}{#1 \delimsize\vert #2}

\DeclareSIUnit\parsec{pc} 
\hypersetup{
    colorlinks=true,
    linkcolor=red,
    filecolor=magenta,      
    citecolor=blue
}

\begin{document}

\title{Constraints on $f(Q)$ logarithmic model using gravitational wave standard sirens}

\author{José Antonio Nájera\orcidlink{0000-0001-9738-7704}}
\email{joseantoniodejesus.najeraquintana@studenti.unipd.it}
\affiliation{Dipartimento di Fisica e Astronomia “Galileo Galilei”, Università degli Studi di Padova\\ via Marzolo 8, I-35131, Padova, Italy}

\author{Carlos Aráoz Alvarado\orcidlink{0000-0002-7095-7874}}
\email{carlosaraoz@ciencias.unam.mx}
\affiliation{Facultad de Ciencias, Universidad Nacional Autónoma de México, Investigación Científica, C.U., Coyoacán, Ciudad de México 04510, México.}

\author{Celia Escamilla-Rivera\orcidlink{0000-0002-8929-250X}}
\email{celia.escamilla@nucleares.unam.mx}
\affiliation{Instituto de Ciencias Nucleares, Universidad Nacional Aut\'{o}noma de M\'{e}xico, 
Circuito Exterior C.U., A.P. 70-543, M\'exico D.F. 04510, M\'{e}xico.}


\begin{abstract}
In this paper, we study the constraints on the $f(Q)=Q/(8\pi G) - \alpha \ln(Q/Q_0)$, symmetric teleparallel model using local measurements and gravitational wave mock standard sirens. Using observational local SNIa and BAO data and energy conditions, the logarithmic $f(Q)$ model is capable of explaining the cosmic late-time acceleration by geometrical means. This result suggests that the logarithmic symmetric teleparallel model could be a candidate to solve the cosmological constant problem. In the case of the simulated standard siren data, by using the performance of the future ET and LISA detectors, we expect to be able to measure the current Hubble constant $H_0$, and the matter content $\Omega_m$, with a precision better than 1\% and 6\%, respectively.
Furthermore, we explore the predicted $f(Q)$ logarithmic model deviation from the standard GR using ET and LISA mock standard sirens. The ratio $d_L^{\text{gw}}(z)/d_L^{\text{em}}(z)$, which quantifies the deviation from GR gives us a significant deviation higher than 13\% at $z=1$, and it continues growing to reach a deviation higher than 18\% in its median value. Future standard siren data will be able to quantify the strength of the deviation from GR and hence whether a cosmology like the one implied by this $f(Q)$ model is feasible. 
\end{abstract}


\maketitle


\section{Introduction}

General Relativity (GR) has been an accomplished theory to describe several cosmological scenarios at different redshifts by being constrained using current observational data within the landscape of the flat $\Lambda$CDM concordance model \cite{Planck2018}. Particularly, under this scheme we can explain the cosmological evolution at different epochs, e.g. the large-scale structure formation (LSS) \cite{Troster2020}, thermal history, and big bang nucleosynthesis (BNN)\cite{Dodelson2021-mi}, the current cosmic accelerated expansion \cite{Sahni2004}, and a wide class of other cosmological observations and measurements \cite{Riess1998,Hajkarim2020}. Moreover, despite its achievements, the concordance model has theoretical and observational issues, e.g. coincidence problem \cite{Weinberg1989}, the fine-tuning problem \cite{Perivolaropoulos2022,Martin2012}, and more recently, a statistical tension between direct and indirect measurements of some characteristic cosmological parameters \cite{DiValentino2021,Abdalla2022}. \\

To alleviate these issues, alternatives beyond the $\Lambda$CDM model in the direction to modify the long-range gravitational interaction as in modified gravity (MG) theories have been proposed \cite{Clifton:2011jh,DiValentino:2021izs} . In this line of thought, the first standard way to derive gravitational modifications is to add extra terms in the Einstein-Hilbert action, which gives a wide range of MG theories e.g $f(R)$ gravity \cite{DeFelice:2010aj}, Horndeski scalar-tensor theories \cite{Horndeski:1974wa}, etc, just to cite a few. A second way is through Teleparallel Gravity (TG), which is a gauge theory where the gravitational field is moderated by the torsion instead of curvature \cite{Aldrovandi:2013wha}. In this TG landscape, we can derive a teleparallel equivalent to GR, so-called TEGR, in which we can study
classical phenomena dynamically equivalent  \cite{Bahamonde:2021gfp,Krssak:2018ywd,Cai:2015emx}. On this matter, every equivalent description of GR using curvature and torsion are some examples to perform modification on gravity theories. Furthermore, a venue in this second way to derive MG theories is through the symmetric teleparallel gravity (STG), based on a function $f(Q)$, where $Q$ is the non-metricity scalar. Recently, this kind of theory has been given interesting cosmological features, e.g. accelerating solutions for inflation and dark energy epochs \cite{BeltranJimenez:2019tme}, dynamical system analysis at background and perturbation linear level \cite{Khyllep:2022spx, Albuquerque:2022eac}, anisotropic $f(Q)$ scenarios \cite{Koussour:2022sab} and $f(Q)$ cosmography at first order \cite{Gadbail:2022hwq}. In this STG terrain it is possible to build an equivalent formulation to GR,  the symmetric teleparallel equivalent to GR, denoted as STEGR \cite{jimenez2018coincident}. These three (GR, TEGR, STEGR) seemingly unrelated
representations of the same underlying theory are known as the geometrical trinity of gravity \cite{beltran2019geometrical, capozziello2022comparing}. Based on these alternative formulations of
GR, we can further our study in their extensions \cite{golovnev2017covariance,najera2022effects}.

On the observational constraint analyses, $f(Q)$ theory has been resulting in cosmic deceleration phases that go into acceleration phases by considering observations from supernovae type Ia (SNIa), Cosmic Microwave Background (CMB), Baryon Acoustic Oscillations (BAO) and $H(z)$ measurements \cite{Lazkoz:2019sjl,Barros:2020bgg,Ayuso:2020dcu,Anagnostopoulos:2021ydo,Atayde:2021pgb,Frusciante:2021sio}. In the light of these observables and their related catalogs, another venue to constrain such theory has been through Gravitational Waves (GW) standard sirens \cite{d2022forecasting}, however, in the analysis discussed, there are not significant deviations with respect to GR. Therefore, one of our goals is to compute the degree of precision in order to notice a significant deviation from this standard GR.

In this work, we present a $f(Q)$ logarithmic gravity model capable of explaining the late-time accelerated expansion of the universe in geometrical terms instead of the use of an exotic fluid with negative pressure. Thus, we will study a model with the potential of solving the cosmological constant problem in the framework of symmetric teleparallel gravity. This can be done by studying the background part and testing whether the model can predict an accelerated expansion. After this, the tensor perturbations part can predict deviations from GR that are expected to observe in a Universe governed by this kind of geometry. The first part is to see whether the model can predict the accelerated expansion behaviour in terms of the deceleration parameter, the energy conditions, and observational constraints from SNIa and BAO data. The second part to study this model is to forecast future standard sirens constraints with the aid of future Einstein Telescope (ET) \cite{maggiore2020science} and LISA \cite{babak2017science} data. By doing so, the deviation from GR can also be derived and hence getting a prediction on how standard siren data should look like in a cosmology like the one implied by this $f(Q)$ modified gravity model. \\

 This paper is divided as follows:
In Sec.~\ref{sec:fQoverview} we present an overview of the $f(Q)$ theory and the cosmological evolution equation under a FLRW symmetry.
In Sec.~\ref{sec:fQmodel} we introduce the $f(Q)$ logarithm model and its evolution equations. Also, we discuss the energy conditions to be fulfilled for a generic $f(Q)$ and the ones that we need to satisfy for the $f(Q)$ logarithm model.
In Sec.~\ref{sec:GW-theory} it is described the form of the gravitational wave luminosity under the coincident gauge.
In Sec.~\ref{sec:datasets}, we describe the observational and simulated baselines used to constrain our $f(Q)$ model.
The results are given In Sec.~\ref{sec:results} and,
finally, we discuss particular insights in Sec.~\ref{sec:conclusions}.


\section{$f(Q)$  gravity scheme}
\label{sec:fQoverview}

We begin by using the Palatini formalism, where two independent objects need to be considered: the metric  $g_{\mu \nu}$ and the connection $\Gamma^\alpha_{\;\;\mu\nu}$, were, greek indices $\mu,\nu, \ldots$ refer to space-time indices. Directly, this relation is given through the field equations of the theory of interest at hand. To start within this scheme, we require to consider both $g_{\mu \nu}$ and $\Gamma^\alpha_{\;\;\mu\nu}$, to establish how we need to perform a parallel transport of an arbitrary tensor along a path in a manifold. 

Furthermore, in this work, we will consider the symmetric teleparallel formulation \cite{adak2006lagrange} where gravity is characterized by the non-metricity tensor $Q_{\alpha \mu \nu} = \nabla_\alpha g_{\mu \nu}$, with the torsion and curvature being equal to zero. The adjective \textit{teleparallel} came from the fact that curvature is zero while symmetric states that the connection is symmetric in its lower indices and hence the torsion is zero. 
According to this, from the non-metricity tensor, we can derive several different useful tensors, e.g. the disformation tensor and the superpotential, respectively \cite{jimenez2020cosmology}
\begin{equation}
    L^\alpha_{\;\;\mu\nu} = - \frac{1}{2}g^{\alpha \beta} (Q_{\nu \mu \beta} + Q_{\mu \beta \nu} - Q_{\beta \mu \nu}),
\end{equation}
\begin{equation}
    P^\alpha_{\;\;\mu\nu} = -\frac{1}{2} L^\alpha_{\;\;\mu\nu} + \frac{1}{4} \left( Q^\alpha - \tilde{Q}^\alpha \right) g_{\mu \nu} - \frac{1}{4} \delta^\alpha_{\;\;(\mu} Q_{\nu)}, 
\end{equation}
where we have defined two traces of the non-metricity tensor as $Q_{\alpha} = g^{\mu \nu} Q_{\alpha \mu \nu}$ and $\tilde{Q}_{\alpha} = g^{\mu \nu} Q_{\nu \alpha \mu}$. With these derived tensors, we can construct the non-metricity scalar \cite{jimenez2020cosmology,jimenez2018coincident}
\begin{equation}
    Q = - Q_{\alpha \mu \nu} P^{\alpha \mu \nu} = -\frac{1}{4} \left( -Q^{\alpha \mu \nu} Q_{\alpha \mu \nu} + 2 Q^{\alpha \mu \nu} Q_{\nu \alpha \mu} -2Q^\alpha \tilde{Q}_{\alpha} + Q^\alpha Q_\alpha  \right).
\end{equation}
Notice that the non-metricity scalar is built from the sum of four contractions of the non-metricity tensor. Additionally, the non-metricity scalar plays the role of the scalar in the Lagrangian as the Ricci scalar in GR. The action associated with $f(Q)$ gravity is given by \cite{jimenez2018coincident}
\begin{equation}
\label{eqn:fQAction}
    S = \int d^4 x \sqrt{-g} \left( - \frac{1}{2} f(Q) + \mathcal{L}_m \right),
\end{equation}
where $f(Q)$ is a function of the non-metricity scalar, $\mathcal{L}_m$ is the matter Lagrangian density and $g$ is the determinant of the
metric $g_{\mu \nu}$. By setting $f(Q) = Q/8\pi G$, we obtain the symmetric teleparallel equivalent of GR (STEGR) \cite{jimenez2018coincident}.  

To compute the non-metricity tensor and, therefore, all the tensors that are defined in terms of it, we need to define a connection. When working in the Symmetric Teleparallel Formulation, we make the following assumptions
\begin{eqnarray}
    R^\alpha_{\;\;\beta\mu\nu} = 0, \quad   T^\alpha_{\;\;\mu\nu} = 0, \quad    Q_{\alpha\mu\nu} \neq 0,
\end{eqnarray}
where $R^\alpha_{\;\;\beta\mu\nu}$ stands for the Riemann tensor and $T^\alpha_{\;\;\mu\nu}$ is the torsion tensor. If we consider curvature equal to zero, we can build a generic connection of the form \cite{hohmann2021variational,beltran2019geometrical}
\begin{equation}
    \Gamma^\mu_{\;\;\nu\beta} = \left( \Lambda^{-1} \right)^\mu_{\;\;\gamma} \partial_\beta \Lambda^\gamma_{\;\;\nu},
\end{equation}
which gives a zero Riemann tensor \cite{hohmann2021variational}. Moreover, since we are considering a torsion equal to zero, we have the additional constraint $\Gamma^\mu_{\;\;[\nu\beta]}=0$. This causes a constraint in the $\Lambda$ tensors $\partial_{[\beta} \Lambda^\gamma_{\;\;\nu]} = 0$. The general solution for this constraint is given by
\begin{equation}
    \Lambda^\gamma_{\;\;\nu} = \partial_\nu \xi^\gamma, 
\end{equation}
which let us write the STG connection as
\begin{equation}
\label{eqn:STGConnection}
    \Gamma^\mu_{\;\;\nu\beta} = \frac{\partial x^\mu}{\partial \xi^\gamma} \partial_\nu \partial_\beta \xi^\gamma,
\end{equation}
where $\partial x^\mu/ \partial \xi^\gamma$ is the inverse of $\partial \xi^\gamma/ \partial x^\mu$. Eq.~ (\ref{eqn:STGConnection}) gives the general form of the connection in STG. In this paper, we will consider the so-called coincident gauge \cite{jimenez2018coincident}. In this gauge, we take the parameters $\xi^\gamma$ as the coordinates and then the connection vanishes. Furthermore, the equations will be easier since the covariant derivatives turn into partial derivatives, however, we lost the diffeomorphism invariance \cite{jimenez2020cosmology}. This issue can be problematic when dealing with cosmological perturbations. By taking the covariant gauge and computing the linear cosmological perturbations, we have already selected the coincident gauge, which makes it impossible to change to the Newtonian or synchronous gauge unless we perform a change of gauge. Varying Eq.~(\ref{eqn:fQAction}) w.r.t the metric and set it to zero ($\delta S = 0$), we obtain the field equations \cite{jimenez2020cosmology}
\begin{equation}
\label{eqn:fieldEquations}
    \frac{2}{\sqrt{-g}} \nabla_\alpha (\sqrt{-g} f_Q P^\alpha_{\;\;\mu\nu}) + \frac{1}{2} f g_{\mu\nu} + f_Q (P_{\mu \alpha \beta} Q_\nu^{\;\;\alpha\beta} - 2Q_{\alpha \beta \mu} P^{\alpha\beta}_{\;\;\;\;\;\nu}) = T_{\mu\nu},
\end{equation}
with $f_Q = df/dQ$. Now, varying the action w.r.t the connection we obtain \cite{jimenez2020cosmology}
\begin{equation}
\label{eqn:connectionEquations}
    \nabla_\mu \nabla_\nu (\sqrt{-g} f_Q P^{\mu\nu}_{\;\;\;\;\;\alpha}) = 0.
\end{equation}

If we raise an index to the field equations (\ref{eqn:fieldEquations}) we get
\begin{equation}
    \frac{2}{\sqrt{-g}} \nabla_\alpha (\sqrt{-g} f_Q P^{\alpha\mu}_{\;\;\;\;\;\nu}) + \frac{1}{2} \delta^\mu_{\;\;\nu} f + f_Q P^{\mu\alpha\beta} Q_{\nu\alpha\beta} = T^\mu_{\;\;\nu}.
\end{equation}
From them, the modified Friedmann equations can be derived under the coincident gauge. We consider the FLRW metric $ds^2 = -dt^2 +a^2(t) dx_i dx_i$ and assume that the Universe is composed of a perfect fluid with density $\rho$ and pressure $p$, then the modified Friedmann equations are given by 
\begin{eqnarray}
    6 f_Q H^2 - \frac{1}{2} f &=& \rho,\label{eqn:firstModifiedFriedmannEquation}\\
       (12H^2 f_{QQ} + f_Q) \dot{H} &=& -\frac{1}{2} (\rho + p).\label{eqn:secondModifiedFriedmannEquation}
\end{eqnarray}

Additionally, by taking the Levi-Civita covariant derivative of the field equations (\ref{eqn:fieldEquations}) and the connection equations (\ref{eqn:connectionEquations}), we get that $\mathcal{D}_\mu T^\mu_{\;\;\nu} = 0$, with $\mathcal{D}$ the Levi-Civita covariant derivative. Then, the standard continuity equation can be read as
\begin{equation}
    \dot{\rho} + 3H(\rho + p) = 0,
\end{equation}
which denotes a conservation equation of the evolution of matter and radiation.

Using the first modified Friedmann equation (\ref{eqn:firstModifiedFriedmannEquation}) in the second modified Friedmann equation (\ref{eqn:secondModifiedFriedmannEquation}) gives the follwoing pressure equation
\begin{equation}
\label{eqn:secondformsecondModifiedFriedmannEquation}
    -2f_Q\left(\dot{H} +3H^2+ 12H^2 \dot{H} \frac{f_{QQ}}{f_{Q}} \right) +\frac{f}{2} =  p.
\end{equation}
In analogy with GR we can rewrite Eq.(\ref{eqn:firstModifiedFriedmannEquation}) and (\ref{eqn:secondformsecondModifiedFriedmannEquation}) as
\begin{eqnarray}
    3H^2&=&\frac{1}{2}\tilde{\rho}, \\
       \dot{H}+ 3H^2&=&-\frac{1}{2}\tilde{p},
\end{eqnarray}
where
\begin{equation}
    \tilde{\rho}=\frac{1}{f_Q}\left(\rho+\frac{f}{2}\right),
     \label{eqn:curlyrho}
\end{equation}

\begin{equation}
    \tilde{p}=24 H^2 \dot{H} \frac{f_{QQ}}{f_{Q}} +\frac{1}{f_Q}\left(p-\frac{f}{2}\right),
       \label{eqn:curlyp}
\end{equation}
where the tilde quantities represent the effective density and pressure. These effective quantities reduce to the standard ones for the STEGR case.

\section{Logarithmic $f(Q)$ model}
\label{sec:fQmodel}

One of the motivations to consider MG theories is to alleviate cosmological issues, for instance, the cosmological constant $\Lambda$ problem. To tackle this problem, it is necessary to find a way to describe the late-time accelerated expansion of the Universe in terms of geometry. In the following analysis, we will test a model with a cosmological constant-like term coming from geometry. We start with a $f(Q)$ function to get the $\Lambda$CDM standard model. Then, we will build our function to predict a cosmic accelerated expansion without a $\Lambda$.

To obtain the Friedmann $\Lambda$CDM equations in $f(Q)$ gravity, we need to consider the following \cite{najera2021fitting}
\begin{equation}
    f(Q) = \frac{Q+2\Lambda}{8 \pi G},
\end{equation}
with $Q/8\pi G$ the STEGR and G as Newton's constant. With this latter function, the first modified Friedmann equation is given by 
\begin{equation}
\label{eqn:LCDMFriedmann}
    H^2 = \frac{\Lambda}{3} + 8\pi G\rho,
\end{equation}
which is precisely the $\Lambda$CDM Friedmann equation. To obtain this result, we need to consider a $f(Q)$ function with the STEGR plus a constant term $\Lambda/(4\pi G)$. This shows that, whenever we take a constant term into the $f(Q)$ function, we are implicitly taking the cosmological constant $\Lambda$. In other words, if we work with a function $f(Q) = f_1(Q) + a$ with $a$ a constant, we are implicitly considering the cosmological constant that can be recovered with the variable change $a = \Lambda/(4\pi G)$.  Therefore, to build a model that can potentially solve the cosmological constant problem, we need to consider a function that does not include a constant term. 

We can now build a $f(Q)$ model that also predicts the cosmic late-time accelerated expansion similar to $\Lambda$CDM if we can get a constant term like the one in the \textit{r.h.s} of Eq.~(\ref{eqn:LCDMFriedmann}). If we look at the first term at \textit{l.h.s} of Eq.~(\ref{eqn:firstModifiedFriedmannEquation}) and we fix it as a constant:
\begin{equation}
    6f_QH^2 = - \alpha,
\end{equation}
and since, in the coincident gauge $Q = 6H^2$, we obtain
\begin{equation}
    \frac{df}{dQ} = - \frac{\alpha}{Q},
\end{equation}
which has the solution
\begin{equation}
\label{eqn:logarithmicTerm}
    f(Q) = - \alpha \ln \left( \frac{Q}{Q_0} \right) + f_0,
\end{equation}
with $Q_0 = Q(z=0)$ and $f_0 = f(z=0)$. However, we will set $f_0 = 0$, since this denotes a constant term. This function will have a constant term on the r.h.s of the first modified Friedmann Eq.~(\ref{eqn:firstModifiedFriedmannEquation}). We can consider this term plus the STEGR to obtain GR when $\alpha = 0$ and a deviation from it when $\alpha \neq 0$. 

\subsection{$f(Q)$ evolution equations}

By considering a logarithmic term in the $f(Q)$ gravity function, we get a cosmological constant-like behaviour in the \textit{l.h.s} of the modified Friedmann equations (\ref{eqn:firstModifiedFriedmannEquation}). This enables us to consider a gravity model with the potential of explaining the cosmic late-time accelerated expansion in terms of geometry. However, since GR has been very well tested in the last 100 years, it is wise to consider it as a starting point and assume a deviation from it. This can be done by minimally coupling the STEGR to the logarithmic term (\ref{eqn:logarithmicTerm}). By doing this, we will start from GR but consider a deviation that can account for the cosmic accelerated expansion. Thus, we will consider the following model
\begin{equation}
    f(Q) = \frac{Q}{8\pi G} - \alpha \ln \left( \frac{Q}{Q_0} \right),
    \label{eqn:f(q)generalmodel}
\end{equation}
which includes the STEGR plus a deviation logarithmic term. The first modified Friedmann Eq. (\ref{eqn:firstModifiedFriedmannEquation}) is now
\begin{equation}
   H^2 + \frac{8\pi G \alpha}{3} \ln \left( \frac{H}{H_0} \right) = \frac{8\pi G\alpha}{3} + \frac{8\pi G\rho}{3}.
\end{equation}
Defining a critical \textbf{geometrical} \textit{density} as $\Omega_\alpha = 8\pi G \alpha / 3H_0^2$, and considering that the universe has a fluid with matter and radiation components we can write
\begin{equation}
\label{eqn:firstFriedmannEquationLogarithmic}
    E^2 + \Omega_\alpha \ln (E) = \Omega_\alpha + \Omega_m (1+z)^3 + \Omega_r (1+z)^4, 
\end{equation}
where 
\begin{eqnarray}
\Omega_m = \dfrac{8\pi G \rho_{0m}}{3H_0^2}, \quad 
\Omega_r = \dfrac{8\pi G \rho_{0r}}{3H_0^2}, \quad
E(z) = \dfrac{H(z)}{H_0}. 
\end{eqnarray}
where the subindex $0$ denotes quantities evaluated at current times. 
Notice that the latter equation cannot be solved analytically, however, a numerical method can compute $E(z)$. Furthermore, by setting $z=0$, we obtain the constraint relation
\begin{equation}\label{eq:constraint}
    \Omega_\alpha + \Omega_m + \Omega_r = 1,
\end{equation}
which reduces the parameter dimension of the model to one parameter (due to this constraint, the logarithmic model has the same parameters as $\Lambda$CDM). Here, $\Omega_\alpha$ can be interpreted as a geometry density, which mimics the $\Lambda$CDM right-hand side term of the Hubble factor for this standard model
\begin{equation}
  E^2_{\Lambda \text{CDM}} = \Omega_\Lambda + \Omega_m (1+z)^3 + \Omega_r (1+z)^4,
\end{equation}
however, it also introduces another term on the left-hand side, which is logarithmic. Therefore, as we can see in Eq. (\ref{eqn:firstFriedmannEquationLogarithmic}) and the constraint (\ref{eq:constraint}), the parameter $\Omega_\alpha = (8\pi G \alpha)/(3H_0^2)$ is behaving similarly to $\Omega_\Lambda$ in $\Lambda$CDM (apart from a logarithmic term which is considerably smaller to $E^2$ and thus it only gives a small contribution in the background framework). However, the parameter $\Omega_\alpha$ comes from the minimal coupling between the STEGR and $\alpha \ln(Q/Q_0)$ instead of coming from the assumption of an exotic fluid with negative pressure as the cosmological constant $\Lambda$. This is the fact that makes this model interesting, the possibility of explaining the late-time accelerated expansion of the Universe in terms of geometrical means. We will confirm this by computing the deceleration parameter (to see if it indeed predicts an accelerated expansion), studying the energy conditions (as a double check to determine the nature of the expansion), and making observational and simulated constraints.  For the second Friedmann equation, we get
\begin{equation}
\label{eqn:secondFriedmannEquationLogarithmic}
    \frac{\dot{H}}{H^2} = - \frac{3 \,\Omega_m (1+z)^3 + 4 \, \Omega_r (1+z)^4}{\Omega_\alpha + 2E^2},
\end{equation}
by using the result from the first Friedmann (\ref{eqn:firstFriedmannEquationLogarithmic}) we can write
\begin{equation}
    \frac{\dot{H}}{H^2} = - \frac{3 \, \Omega_m (1+z)^3 + 4 \, \Omega_r (1+z)^4}{\Omega_\alpha (3 - 2 \ln E) + 2 \, \Omega_m (1+z)^3 + 2 \, \Omega_r (1+z)^4}.
\end{equation}

\begin{figure*}
    \centering
    \includegraphics[scale=0.6]{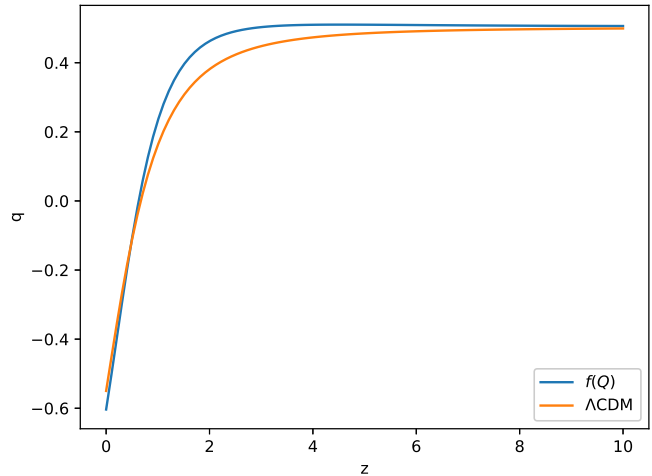}
    \caption{Comparison of the deceleration functions $q(z)$ for the logarithmic $f(Q)$ model (orange color curve) and $\Lambda$CDM (blue color curve). We consider the benchmark case with $\Omega_m = 0.3$, $\Omega_\alpha = \Omega_\Lambda = 0.7$.}
    \label{fig:decelerationParameter-comparison}
\end{figure*}

In this case, the deviation from the $\Lambda$CDM model is more significant 
\begin{equation}
    \left( \frac{\dot{H}}{H^2} \right)_{\Lambda \text{CDM}} = - \frac{3 \, \Omega_m (1+z)^3 + 4 \, \Omega_r (1+z)^4 }{2 \, \Omega_\Lambda + 2 \, \Omega_m (1+z)^3 + 2 \, \Omega_r (1+z)^4},
\end{equation}
where the deviation is given by the $\Omega_\alpha$ term. For the logarithmic $f(Q)$ model, the deviation is given by the function $3/2 - \ln{E}$, while it is a constant for $\Lambda$CDM. From this result, we can determine the deceleration function $q(z) = -(\dot{H}/H^2 + 1) $,
\begin{equation}
    q(z) = \frac{\dfrac{1}{2}\,\Omega_m (1+z)^3 + \Omega_r (1+z)^4 - \Omega_\alpha \left( \dfrac{3}{2} - \ln E \right)}{\Omega_m (1+z)^3 + \Omega_r (1+z)^4 + \Omega_\alpha \left( \dfrac{3}{2} - \ln E \right)},
\end{equation}
and the deceleration parameter evaluated at $z=0$ is
\begin{equation}
\label{eqn:decelParam}
    q_0 = \frac{4\,\Omega_m + 5\Omega_r - 3}{3 - \Omega_M - \Omega_r}.
\end{equation}
For $\Lambda$CDM, it is given by 
\begin{equation}
    q_{\Lambda \text{CDM}}(z) = \frac{\dfrac{1}{2} \, \Omega_m (1+z)^3 + \Omega_r (1+z)^4 - \Omega_\Lambda}{\Omega_m (1+z)^3 + \Omega_r (1+z)^4 + \Omega_\Lambda},
\end{equation}
and
\begin{equation}
    q_{0 \, \Lambda\text{CDM}} = \frac{3}{2} \Omega_m + 2 \Omega_r - 1.
\end{equation}
If we consider the benchmark model with $\Omega_m = 0.3$ and negligible radiation ($\Omega_r \approx 0$), then $q_0 = -2/3$ and $q_{0 \, \Lambda \text{CDM}} = -0.55$. Thus, in the benchmark model, the universe has higher acceleration at present times.

In Figure \ref{fig:decelerationParameter-comparison} we present the comparison between the deceleration function of the logarithmic $f(Q)$ model concerning $\Lambda$CDM in the benchmark case with $\Omega_m = 0.3$ and thus $\Omega_\alpha = 0.7$ for $f(Q)$ and $\Omega_\Lambda = 0.7$ for $\Lambda$CDM. Notice that at high redshift, both curves are nearby each other and they start to differ in the late cosmic times. In particular, the deceleration function starts to decrease earlier in $\Lambda$CDM than in the $f(Q)$ model. However, once the logarithmic model starts to decrease the decreasing rate of the $f(Q)$ function is considerably greater than the one from the $\Lambda$CDM model. Finally, we can see that at present time, the deceleration parameter is smaller for the $f(Q)$ model, indicating a bigger cosmic late-time expansion. In the range $z \in [0,2]$, we have a point $z_c$ where $q_{f(Q)}(z_c) = q_{\Lambda \text{CDM}}(z_c)$. An additional feature that we can spot in figure \ref{fig:decelerationParameter-comparison} is the fact that the $f(Q)$ gravity model is indeed predicting a late-time accelerated expansion.


\subsection{Energy Conditions discussion}

As a second proof that the model is predicting an accelerated expansion of the Universe, we will work the energy conditions. With the model presented in Eq.(\ref{eqn:f(q)generalmodel}), we can now derive the strong, weak, null, and dominant energy conditions. The energy conditions (ECs) are tools that grant the geodesic and causal structure of space-time. These ECs can be derived from the Raychaudhuri Eq.~(\ref{eqn:secondFriedmannEquationLogarithmic}), which gives the behaviour of attractiveness and congruence of gravity for null, timelike, or lightlike curves, respectively \cite{arora2021constraining}. In this work, we consider the timelike and null curves, whose forms are, respectively 
\begin{align}
    & R_{\mu \nu}U^\mu U^\nu+\frac{1}{3}\theta^2+\sigma_{\mu \nu}\sigma^{\mu \nu}-\omega_{\mu \nu}\omega^{\mu \nu}+\frac{d\theta}{d\tau}=0, \label{eqn:timelikeRaychaudhuriequation} \\
    & R_{\mu \nu}n^\mu n^\nu+\frac{1}{2}\theta^2+\sigma_{\mu \nu}\sigma^{\mu \nu}-\omega_{\mu \nu}\omega^{\mu \nu}+\frac{d\theta}{d\lambda}=0, \label{eqn:spacelikeRaychaudhuriequation}
\end{align}
where $\theta$ is the expansion factor which stands for the expansion of volume, $\sigma_{ \mu \nu}$ is the shear tensor that measures the distortion of the volume, $\omega_{ \mu \nu}$ is the vorticity or rotation tensor which measures the rotation of the curves, $U^\mu$ is the timelike vector and $n^\mu$ is the lightlike vector; both tangents to the curves, and  $\tau$ and $\lambda$ are positive parameters used to describe the curved of the congruence
\cite{alvarenga2012testing}. Equations (\ref{eqn:timelikeRaychaudhuriequation}) and (\ref{eqn:spacelikeRaychaudhuriequation}) satisfy the following conditions for attractive gravity
\begin{align}
    & R_{\mu \nu}U^\mu U^\nu \geq 0, \\
    & R_{\mu \nu}n^\mu n^\nu \geq 0.
\end{align}
Consequently, as we are considering a perfect fluid matter distribution, the energy conditions that are recovered from GR are given by \cite{mandal2020energy}
\begin{itemize}
    \item Strong Energy Conditions (\textbf{SEC}) if $\tilde{\rho} + 3\tilde{p}\geq 0 $,
    \item Weak Energy Conditions (\textbf{WEC})  if $\tilde{\rho}\geq 0 $, $\tilde{\rho} + \tilde{p}\geq 0 $,
    \item Null Energy Conditions (\textbf{NEC})  if $\tilde{\rho} + \tilde{p}\geq 0 $,
    \item Dominant Energy Conditions (\textbf{DEC})  if $\tilde{\rho} \geq 0 $, $|\tilde{p}| \leq \tilde{\rho} $.
\end{itemize}
Taking Eqs.~(\ref{eqn:curlyrho})-(\ref{eqn:curlyp}) into the above ECs constraints we obtain the following:
\begin{enumerate}
\item For the \textbf{SEC} case we obtain
\begin{equation}
   \rho+3p-f+72 H^2 \dot{H} f_{QQ} \geq 0
\end{equation}

\item For the \textbf{WEC} case we arrive to the constraints
\begin{equation}
  \rho+ \frac{f}{2} \geq 0,  \kern 2.5pc \rho + p +24\dot{H}H^2 f_{QQ} \geq 0.
  \end{equation}

\item For the \textbf{NEC} case we obtain
\begin{align}
      \rho + p +24\dot{H}H^2 f_{QQ}\geq 0
\end{align}

\item For the \textbf{DEC} case we obtain
\begin{equation}
     \rho +\frac{f}{2} \geq 0
\end{equation}

\end{enumerate}

And for the constraint $|\tilde{p}| \leq \tilde{\rho} $ as $|\tilde{p}| \leq \tilde{\rho} \iff -\tilde{\rho} \leq \tilde{p} \leq \tilde{\rho}$, we obtain
\begin{equation}
   -\rho -24 \dot{H}H^2 f_{QQ} \leq p \leq \rho +f -24 \dot{H}H^2  f_{QQ}
\end{equation}

Let us consider the cosmological kinematic parameters described by
 \begin{align}
     & H=\frac{\dot{a}}{a},  \qquad \qquad
      q=-\frac{1}{H^2}\frac{\Ddot{a}}{a},   \\
     & j=\frac{1}{H^3}\frac{\dddot{a}}{a}, \qquad  \qquad s=\frac{1}{H^4}\frac{\ddddot{a}}{a}, \nonumber
      \end{align}
where $H, q, j, s$ are the Hubble, deceleration, jerk, and snap parameters, respectively. $a(t)$ is the scale
factor in the FLRW metric. Such cosmological kinematic parameters allow us to represent the time derivatives of $H$ as
 \begin{equation}
     \dot{H}=-H^2(1+q)
    \label{eqn:Hp}
 \end{equation}
  \begin{equation}
     \Ddot{H}=-H^3(j+3q+2)
 \end{equation}
By using Eq.(\ref{eqn:Hp}) we can rewrite Eqs.~(\ref{eqn:firstModifiedFriedmannEquation})-(\ref{eqn:secondformsecondModifiedFriedmannEquation}) as
\begin{equation}
    \rho=6f_QH^2-\frac{f}{2},
    \label{eqn:densityq}
\end{equation}
\begin{equation}
   p= -2f_Q\left[2H^2-H^2q -12H^4(1+q) \frac{f_{QQ}}{f_{Q}} \right] +\frac{f}{2},
   \label{eqn:pressureq}
\end{equation}
which correspond to the density and pressure of the $f(Q)$ gravity.

Therefore, the density (\ref{eqn:densityq}) and pressure (\ref{eqn:pressureq}) of the $f(Q)$ gravity establish the
following constraints for the energy conditions
\begin{itemize}
\item \textbf{SEC}: For $  \rho+3p-f+72H^2\dot{H}f_{QQ} \geq 0$ we obtain
\begin{equation}
    -f_QH^2+f_QH^2q \geq 0
\end{equation}

\item \textbf{WEC}: For $  \rho+p+24H^2\dot{H}f_{QQ} \geq 0$ we obtain
\begin{equation}
    f_QH^2+f_QH^2q \geq 0
\end{equation}
and for $\rho+\frac{f}{2} \geq 0$ we obtain
\begin{equation}
    f_{Q}H^2 \geq 0
\end{equation}

\item  \textbf{NEC}: For $  \rho+p+24H^2\dot{H}f_{QQ} \geq 0$ we obtain
\begin{equation}
    f_QH^2+f_QH^2q \geq 0
\end{equation}

\item \textbf{DEC}:  For $  \rho+\frac{f}{2} \geq 0$ we obtain
\begin{equation}
    f_{Q}H^2  \geq 0
\end{equation}
and for 
$
   -\rho -24 \dot{H}H^2 f_{QQ} \leq p \leq \rho +f -24 \dot{H}H^2  f_{QQ}
$ we obtain
\begin{equation}
    -3f_QH^2 \leq -2f_QH^2+f_Q H^2q\leq 3f_QH^2.
\end{equation}
\end{itemize}

To discuss the viability of the functional form of $f(Q)$ in FLRW spacetime, we test which ECs are fulfilled by the $f(Q)$ logarithmic model.
We use the general model of the functional form of $f(Q)$ (\ref{eqn:f(q)generalmodel}) to analyze the ECs with the current values of the geometric parameters in $f(Q)$ theory. The previous functions establish that the ECs need to satisfy the following constraints:
 \begin{eqnarray}
     \text{For \textbf{SEC}:}\quad && \left[\frac{1}{8\pi G}- \frac{\alpha}{6H_0^2}\right]H_0^2(q_0-1) \geq 0, \\
         \text{For \textbf{WEC}:}\quad &&  \left[\frac{1}{8\pi G}- \frac{\alpha}{6H_0^2}\right]H_0^2(1+q_0) \geq 0, \\
          \text{and}\quad  &&  \left[\frac{1}{8\pi G}- \frac{\alpha}{6H_0^2}\right]H_0^2 \geq 0,
 \end{eqnarray}
 
 \begin{eqnarray}
     \text{For \textbf{NEC}:}\quad && \left[\frac{1}{8\pi G}- \frac{\alpha}{6H_0^2}\right]H_0^2(1+q_0) \geq 0, \\
      \text{For \textbf{DEC}:}\quad  &&  \left[\frac{1}{8\pi G}- \frac{\alpha}{6H_0^2}\right]H_0^2 \geq 0,
 \end{eqnarray}
 
  \begin{equation}
  \text{and} \quad  3 \left[\frac{1}{8\pi G}- \frac{\alpha}{6H_0^2}\right]H_0^2 \pm \left[\left[\frac{1}{8\pi G}- \frac{\alpha}{6H_0^2}\right]H_0^2(q_0-2)\right] \geq 0.
 \end{equation}

By plugging the definition $\Omega_\alpha = (8\pi G \alpha)/(3H_0^2)$, using the constraint relation (\ref{eq:constraint}) and the value of the deceleration parameter (\ref{eqn:decelParam}) we get the constraints
 \begin{enumerate}
     \item \textbf{SEC:} $\dfrac{6}{5} \leq \Omega_m < 3$ 
     \item \textbf{WEC:} $-1\leq \Omega_m < 3$
     \item \textbf{NEC:} $\Omega_m < 3$
     \item \textbf{DEC:} $0 \leq \Omega_m \leq 2$
 \end{enumerate}

Notice that the energy conditions depend on $\Omega_m$ solely. For the physically expected values of this parameter $\Omega_m \in [0,1]$, all the energy conditions are full-filled except SEC. Therefore, in addition to the deceleration function, the fulfillment of WEC, NEC, and DEC, and the violation of SEC corroborates an accelerated expansion \cite{visser2000energy,mandal2020energy}.

\section{Gravitational wave luminosity distance in the STG coincident gauge}
\label{sec:GW-theory}

By fixing the coincident gauge, we have lost the freedom to choose another one when dealing with perturbation theory. We would need either to start from a generic gauge and work with the most general connection \cite{d2021revisiting,hohmann2021general} or derive the perturbation equations in the coincident gauge and then perform a gauge transformation. Fortunately, the transverse tensor component is gauge invariant \cite{piattella2018lecture}. 
By breaking diffeomorphisms nondynamical backgrounds can appear in this scheme, where the standard relations between dynamics and geometric quantities must secure that there are consistent before ruling them out. In addition to this, we can see that by changing gauge, the form of the transverse trace-less tensor perturbation is unchanged. This can be seen by performing a gauge change with the infinitesimal coordinate transformation $\tilde{x}^\mu = x^\mu + \xi^\mu(x)$. As we can see, this transformation represents a local symmetry (it can have different values in different points of space). Therefore, a rank-2 tensor cannot realise the local nature of this symmetry because if it could, it would violate the properties of a rank-2 tensor. This can be written as $\tilde{h}_{ij} = h_{ij}$.

In linear perturbation theory, the evolution of the transverse tensor part is given by \cite{jimenez2020cosmology}
\begin{equation}
\label{eqn:transverseTensor}
    h''_\lambda + 2\mathcal{H} \left[ 1 + \frac{1}{2\mathcal{H}} \frac{d\ln f_Q}{d\eta} \right] h'_\lambda + k^2 h_\lambda = 0,
\end{equation}
which describes the propagation of GWs in the $f(Q)$ theory, where $\eta$ is the conformal time, $\mathcal{H}$ is the conformal Hubble factor, and $\lambda = \times, +$, labels the two GWs polarizations. However, even though $h_{ij}$ is gauge invariant, the Hubble factor is not since the modified Friedmann equations depend on the connection that is being used \cite{d2021revisiting,hohmann2021general}. Thus, the results presented in this paper are valid for the coincident gauge and how GWs propagate with a different connection constitutes an interesting topic for future research. In general, theories that follow the relation
\begin{equation}
    h''_\lambda + 2\mathcal{H} [1 - \delta(\eta)] h_\lambda + k^2 h_\lambda = 0,
\end{equation}
where $\delta(\eta)$ is known as the friction term, the GW luminosity distance is determined by \cite{belgacem2019testing}
\begin{equation}
    d_L^{\text{gw}}(z) = d_L^\text{em}(z) \exp \left[- \int_0^z dz' \frac{\delta(z')}{1+z'}\right],
\end{equation}
with $d_L^{\text{gw}}(z)$ the GW luminosity distance, $d_L^\text{em}(z)$ the standard electromagnetic luminosity distance and $\delta(z)$ the friction term. Also, we can compute the GW luminosity distance with the friction term through the following expression
\begin{equation}
    \delta(\eta) = -\frac{1}{2\mathcal{H}} \frac{d\ln f_Q}{d\eta}.
\end{equation}

The meaning of the friction term is the fact that when GWs travel around space-time, their amplitude of them decays as $1/(a\sqrt{f_Q})$ \cite{jimenez2020cosmology} instead of $1/a$ as in GR. By quantifying deviations of the GW luminosity distance concerning the electromagnetic luminosity distance, we would be able to measure deviations from GR. Straightforwardly, we will derive the theoretical expression to account for the deviation of these two distances. We begin by considering
\begin{equation}
    \delta(\eta) = -\frac{1}{2\mathcal{H}} \frac{d\ln f_Q}{d\eta} = \frac{d\ln \sqrt{f_Q}}{d\ln(1+z)},
\end{equation}
and then
\begin{equation}
    \delta(z) = -\frac{\Omega_\alpha \dot{H}}{(2E^2 -\Omega_\alpha) H^2},
\end{equation}
and taking the ratio of $\dot{H}/H^2$ from Eq.~(\ref{eqn:secondFriedmannEquationLogarithmic})
\begin{equation}
    \delta(z) = \Omega_\alpha \, \frac{3\Omega_M(1+z)^3 + 4\Omega_r (1+z)^4}{4E^4(z)-\Omega_\alpha^2},
\end{equation}
therefore, the GW luminosity distance for this logarithmic $f(Q)$ model is
\begin{equation}
\label{eqn:standardSiren}
    d_L^\text{gw}(z) = d_L^\text{em}(z) \exp \left(- \Omega_\alpha \int_0^z dz' \;\frac{3\,\Omega_M (1+z')^2 + 4\,\Omega_r (1+z')^3}{4E^4(z') - \Omega^2_\alpha} \right).
\end{equation}

From Eq.~(\ref{eq:constraint}) we can notice that the only way to satisfy that $d_L^\text{gw}(z) = d_L^\text{em}(z)$, for all redshift is when $\Omega_\alpha = 0$. However, from the model we are considering, this would constitute a Universe filled solely with matter and radiation in contradiction with observations. To describe the accelerated late-time expansion of the universe, we need the parameter $\Omega_\alpha$ to be different from zero and also to play the role of a dark energy component. Hence, to account for this observational fact, we will have deviations from the standard luminosity distance for all real values of $\Omega_\alpha$. We can also see that even though the $f(Q)$ model in the background framework has solutions similar to $\Lambda$CDM, in the perturbations regime they are much more significant by modifying the amplitude of GWs when traveling in the Universe. We will compute the strength of that deviation with the aid of simulated GW baselines.

\section{Observational and simulated baselines}
\label{sec:datasets}

In this section, we describe the observational and simulated data used to constrain our present $f(Q)$ logarithmic model. For this, we are going to consider a statistical analysis combining the simulated catalogs from GW as the simulated baseline and a set of two observational catalogs as part of the observational baseline.
The set of best fits for our constrained parameters of interest can be obtained through a process with a modified version on \texttt{MontePython}\footnote{\href{https://monte-python.readthedocs.io/en/latest/}{monte-python.readthedocs.io}} and \texttt{CLASS}\footnote{\href{http://class-code.net}{class-code.net}} for our cosmology and both of baselines and the extract of constraints using the plotting tool of \texttt{MontePython}. We used the Metropolis-Hastings method with enough chains to satisfy the Gelman-Rubin convergence criterion $R-1<0.03$ \cite{gelman1992inference}. For the considered $f(Q)$ logarithmic gravity model, the parameter vector is given by $\Theta = \{ H_0, \Omega_m
 \}$ (the same dimension as $\Lambda$CDM) and the derived parameter is given by $\Omega_\alpha$, which is a parameter causing the accelerated expansion of the Universe via a minimal coupling to the STEGR. We begin by describing the observational baselines to be used. From this data, we will derive an observational value of $\Omega_m$ that we will use later to generate the simulated baselines. We will use observational local baselines to study the $f(Q)$ model at a background perspective and thus making the third check that it can predict a late-time accelerated expansion of the Universe. After that, we will study the deviation from GR with simulated standard sirens. This will allow us to forecast the difference between GR and the $f(Q)$ model.

\subsection{Observational baselines}

\begin{itemize}
\item \textbf{Pantheon sample from Type Ia Supernovae (SNIa)} - Type Ia supernovae observables are the first data that probed the cosmic acceleration at late times. This is related to an exotic fluid-like component of negative pressure, a.k.a dark energy. In this work, we use the Pantheon compilation of the SNIa data comprising 1048 data points spanned in the redshift interval $[0.01, 2.3]$~\cite{Pan-STARRS1:2017jku}. At redshift $z_i$, the distance modulus is computed through
\begin{equation}
    \mu(z_i, \Theta) = m - M = 5 \log_{10}[D_L(z_i, \Theta)] + 25 \,,
\end{equation}
where $D_L(z_i, \Theta)$ is the luminosity distance given by
\begin{equation}
    D_L(z_i, \Theta) = c(1+z_i) \int_0^{z_i} \frac{dz'}{H(z', \Theta)} \,. 
\end{equation}
Additionally, the apparent magnitude of each SNIa should be calibrated via an arbitrary fiducial absolute magnitude $M$, and in the MCMC analyses, we can treat $M$ as a nuisance parameter by marginalizing it. The cosmological parameters are then constrained by minimizing a $\chi^2$ with
\begin{equation}
    \chi^2_{\mathrm{SN}} = (\Delta \mu(z_i), \Theta))^T C^{-1} (\Delta \mu(z_i), \Theta)) \,,
\end{equation}
where $(\Delta \mu(z_i), \Theta)) = (\mu(z_i), \Theta) - \mu(z_i)_{\mathrm{obs}}$ and $C$ is the covariance matrix that accounts for the statistical and systematic uncertainties. 

\item \textbf{Baryon Acoustic Oscillations (BAO)} - We also consider the baryon acoustic oscillation data set consisting of independent data points. This includes measurements from the SDSS Main Galaxy Sample at $z_{\mathrm{eff}} = 0.15$ \cite{Ross:2014qpa}, the six-degree Field Galaxy Survey at $z_{\mathrm{eff}} = 0.106$ \cite{2011MNRAS.416.3017B}, and the BOSS DR11 quasar Lyman-alpha measurement at $z_{\mathrm{eff}} = 2.4$ \cite{Bourboux:2017cbm}. 
Addtionally, we consider the angular diameter distances and $H(z)$ measurements of the SDSS-IV eBOSS DR14 quasar survey at $z_{\mathrm{eff}} = \{0.98, 1.23, 1.52, 1.94\}$ \cite{Zhao:2018gvb}, along with the SDSS-III BOSS DR12 at $z_{\mathrm{eff}} = \{0.38, 0.51, 0.61\}$ \cite{Alam:2016hwk}. 
For these datasets we compute the Hubble distance $D_H(z)$, comoving angular diameter distance $D_M(z)$, and volume average distance $D_V(z)$ through:
\begin{equation}
  D_V(z) = \left[(1+z)^2D_A(z)^2 \frac{c z}{H(z)}\right]^{1/3}
\end{equation}
where $D_A(z)=(1+z)^{-2}D_L(z)$ is the angular diameter distance. Afterwards, we compute the combination of parameters $\mathcal{G}(z_i)=D_V(z_i)/r_s(z_d),\allowbreak\,r_s(z_d)/D_V(z_i),\allowbreak\,D_H(z_i),\allowbreak\,D_M(z_i)(r_{s,\mathrm{fid}}(z_d)/r_s(z_d)),\allowbreak\,H(z_i)(r_s(z_d)/r_{s,\mathrm{fid}}(z_d)),\allowbreak\,D_A(z_i)(r_{s,\mathrm{fid}}(z_d)/r_s(z_d))$, 
for which the comoving sound horizon at the end of the baryon drag epoch at redshift $z_d\approx1059.94$ \cite{Planck:2018vyg} is given by:
\begin{equation}
    r_s(z)=\int_z^\infty\frac{c_s(\tilde{z})}{H(\tilde{z})}\,\mathrm{d}z=\frac{1}{\sqrt{3}}\int_0^{1/(1+z)}\frac{\mathrm{d}a}{a^2H(a)\sqrt{1+\left[3\Omega_{b,0}/(4\Omega_{\gamma,0})\right]a}}\,,
\end{equation}
where we have adopted $\Omega_{b,0}=0.02242$ \cite{Planck:2018vyg}, $T_{0}=2.7255\,\mathrm{K}$ \cite{2009ApJ...707..916F}, and a fiducial value of $r_{s,\mathrm{fid}}(z_d)=147.78\,\mathrm{Mpc}$.
Finally, the $\chi^2$ for the BAO data is
\begin{equation}
\chi^2_{\text{BAO}}(\Theta) = \Delta G(z_i,\Theta)^T C_{\text{BAO}}^{-1}\Delta G(z_i,\Theta),
\end{equation}
where $\Delta G(z_i,\Theta) = G(z_i,\Theta)-G_{\text{obs}}(z_i)$and $C_{\text{BAO}}$ is the covariance matrix of all these data.
\end{itemize}

\subsection{Simulated baselines}

We generate mock future GW catalogs. Particularly, we focus on two future GW detectors: the Einstein Telescope (ET) \footnote{\href{https://www.et-gw.eu}{et-gw.eu}} and the Laser Interferometer Space Antenna (LISA)\footnote{\href{https://lisa.nasa.gov}{lisa.nasa.gov}}. The method to generate these events is described in \cite{cai2018probing,d2019probing}. Furthermore, this method has been used in \cite{d2022forecasting} to constrain an $f(Q)$ power law with mock GW catalogs. 
\begin{itemize}
\item \textbf{Einstein Telescope survey}.-
The ET is a planned European ground-based gravitational wave detector that will work on frequencies from 1 Hz up to $10^4$ Hz \cite{d2022forecasting}. ET will be able to measure GWs coming from the collision of binary black holes up to redshift $z \sim 20$ with a rate of $10^5-10^6$ events per year \cite{maggiore2020science}. In addition to this, it will be capable of detecting GWs coming from the collision of binary neutron stars up to redshift $z \sim 2-3$ with a rate of the order of $10^5$ events per year \cite{maggiore2020science}. From these events, after about ten years of operation, it will be possible to identify around 1000 standard sirens up to redshift $z \sim 5$ \cite{cai2018probing, d2022forecasting}. 
The purpose is to generate a mock standard sirens catalog matching this prediction, i.e., a 1000 standard sirens catalogue generated by taking into account the logarithmic cosmological model and the sensitivity of the ET. For this, we need to generate the redshifts, gravitational wave luminosity distances, and their uncertainties. The mock redshift depends on both the cosmological model and the compact object merger rate. We consider the rates given in \cite{cai2018probing,d2019probing}. The next step is to generate the fiducial value of $d_{L \, \text{fiducial}}^{\text{gw}}(z)$, which is generated by assuming a certain cosmology. In this work, we consider our logarithmic $f(Q)$ model. For the cosmological parameters, we have
\begin{itemize}
    \item $\mathbf{H_0:}$  we consider the mean value of the most recent result from the SH0ES collaboration of $H_0 = 73.04 \, \pm \, 1.04$ km/s/Mpc \cite{riess2022comprehensive}. 
    \item $\mathbf{\Omega_m:}$ We consider the mean value of the results of our MCMC assuming the $f(Q)$ model and the full observational baseline (Pantheon+BAO). 
\end{itemize}
For the uncertainties $\sigma_{d_L^{\text{gw}}}$, we computed them following the procedure of \cite{cai2018probing} where the signal-to-noise ratios need to be derived. After generating this errors, we compute the final $d_L^{\text{gw}}$ by using the expression
\begin{equation}
\label{eqn:mockdlGW}
    d_{L \, \text{mock}}^{\text{gw}}(z) \sim \mathcal{N}(d_{L \, \text{fiducial}}^{\text{gw}} (z), \sigma_{d_L^{\text{gw}}}^2).,
\end{equation}
and for this case
\begin{equation}
\label{eqn:chi2GW}
    \chi^2_{GW} = \sum_{i=1}^N \left( \frac{d_L^{gw}(z_i, \Theta) - d_{L \, \text{mock}}^{gw}(z_i)}{\sigma_{d_{L \, \text{mock}}^{gw}}(z_i)}
 \right)^2,
\end{equation}
with $d_L^{gw}(z_i,\Theta)$ the theoretical GW luminosity distance from Eq. (\ref{eqn:standardSiren}).

\item \textbf{LISA survey}.-
LISA is a future planned space GW detector that will work in frequencies from $10^{-4}$ Hz to $2.5 \, \times \, 10^{-2}$ Hz \cite{cai2018probing}. This detector will be composed of three spacecraft components that will form an equilateral triangle. The distance between the two of them is going to be or the order $2.5$ Gm \cite{babak2017science}. LISA will be capable of detecting GWs coming from the collision of massive black hole binaries (MBHB) with masses between $10^4 \, M_\odot$ and $10^7 \, M_\odot$ \cite{babak2017science}. This kind of collision is the most likely that will have an associated electromagnetic counterpart and hence, they have the highest chance to be standard sirens \cite{d2022forecasting}. 
We considered three kinds of astrophysics models for the MBHB sources
\begin{itemize}
    \item \textbf{Pop III:} This is a light-seed model where MBHB forms from the remnants of population III stars at redshift $z \sim 20$ \cite{barausse2012evolution}. 
    \item \textbf{Delay:} This is a heavy-seed model where MBHB is formed from the collapse of protogalactic disks  \cite{tamanini2016science}. This model includes delays.
    \item \textbf{No delay:} The same as the previous one but without including delays. This model is considered as an optimistic upper bound scenario for eLISA \cite{tamanini2016science}.
\end{itemize}
In this case, for the redshift distribution, we consider the results from \cite{tamanini2017late} where the number of events for the Pop III model is 28, for Delay 27, and for No delay 41. This number of events is a representative expected value for a five-year operation of LISA. Hence, we built a catalog with 96 mock standard sirens for this future detector. In addition, we generated the fiducial values of $d_L^{\text{gw}}$ assuming our logarithmic $f(Q)$ model and the same cosmological parameters as in the generated ET mock catalogue. Moreover, as in the previous case, we generated the uncertainties with the method presented in \cite{cai2018probing} and the mock gravitational wave luminosity distance from the normal distribution (\ref{eqn:mockdlGW}). Finally, to perform the MCMC procedure we use Eq. (\ref{eqn:chi2GW}).
\end{itemize}

\begin{figure*}
    \centering
    \includegraphics[scale=0.54]{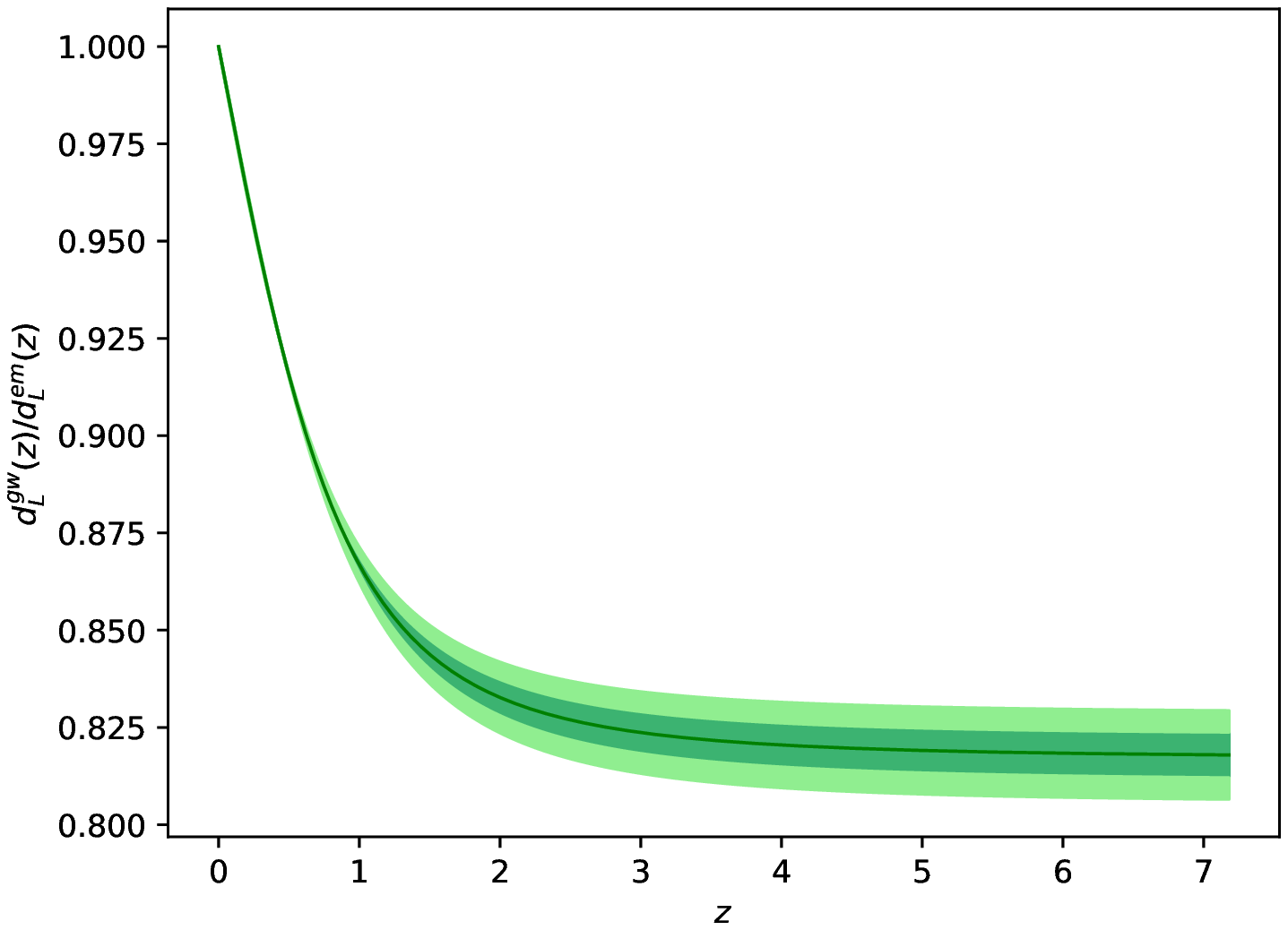}
    \includegraphics[scale=0.54]{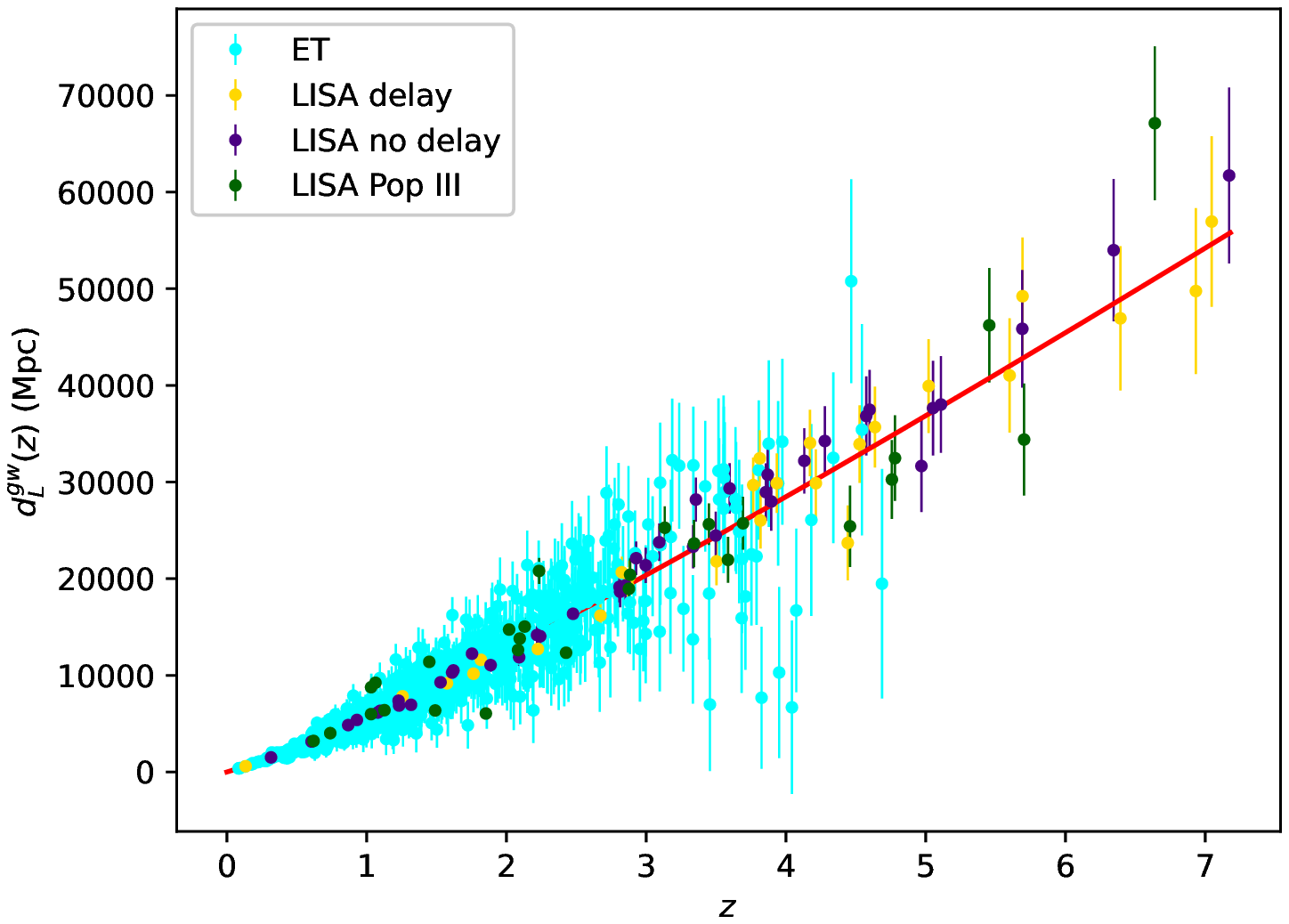}
    \caption{\textit{Left:} Mean value (dark green) and C.L. at 1-$\sigma$ (medium green) and 2-s$\sigma$ (light green) for the ratio of the gravitational wave luminosity distance concerning the standard candle luminosity distance. We consider the mean and uncertainties for the joint ET+LISA catalog. \textit{Right:} Simulated standard sirens for the ET and LISA with their uncertainties at 1-$\sigma$. The red curve represents the $d_L^{gw}(z)$ evolution for the mean value of the cosmological parameters assuming the $f(Q)$ logarithmic model and the mock baseline ET+LISA.}
    \label{fig:logarithmicRatio}
\end{figure*}

Finally, in Figure \ref{fig:logarithmicRatio} we present the evolution of the ratio $d_L^{\text{gw}}(z)/d_L^{\text{em}}(z)$ with redshift. We included the mean value along with the C.L. at 1-$\sigma$ and 2-$\sigma$. In this figure, we also included a plot of the simulated baseline ET and LISA plus the best fit computed from the mean values and the joint ET+LISA baseline. 

\section{Analyses}
\label{sec:results}

We worked with uniform prior probabilities for the cosmological parameters $H_0 \sim \mathcal{U}(0,100)$ km/s/Mpc, and $\Omega_m \sim \mathcal{U}(0,1)$. 
The mean values along with the 1-$\sigma$ and 2-$\sigma$ uncertainties for the cosmological parameters $H_0, \Omega_m$ and the derived parameter $\Omega_\alpha$ are reported in Table \ref{tab:mean_values}. We repeated the MCMC process for the joint observational baseline Pantheon+BAO, the individual simulated baselines ET and LISA, the joint simulated baseline ET+LISA and the joint baseline Pantheon+BAO+ET+LISA containing both observational and simulated data. \\

\begin{table}[h]
    \centering
    \begin{tabular}{|c|c|c|c|c|}
    \hline 
       Database & $H_0$ (km/s/Mpc) & $\Omega_m$ & $\Omega_\alpha$ \\ \hline
       Pantheon+BAO & - & $0.2993^{+0.0052 (+0.0118)}_{-0.0061(-0.0110)} $& $0.7005^{+0.0061 (+0.0110)}_{-0.0052 (-0.0118)}$ \\ \hline
       ET & $72.66^{+0.74(+1.28)}_{-0.64 (-1.36)}$ & $0.3200^{+0.0215 (+0.0505)}_{-0.0276 (-0.0463)}$ & $0.6798^{+0.0276(+0.0463)}_{-0.0215 (-0.0505)}$ \\ 
       \hline
       LISA & $72.16_{-0.87(-1.73)}^{+0.87(+1.73)}$ & $0.3573_{-0.0348(-0.0612)}^{+0.0281 (+0.0642)}$ & $0.6426_{-0.0281 (-0.0642)}^{+0.0348 (+0.0612)}$ \\ \hline 
       ET+LISA & $72.48_{-0.51(-1.02)}^{+0.51(+1.04)}$ & $0.3355_{-0.0194(-0.0369)}^{+0.0173(+0.0375)}$ & $0.6644_{-0.0173(-0.0375)}^{+0.0194(+0.0369)}$ \\ \hline 
       Pantheon+BAO+ET+LISA & $73.50_{-0.28(-0.55)}^{+0.27(+0.55)}$ & $0.2988_{-0.0057(-0.0112)}^{+0.0055(+0.0113)}$ & $0.7012_{-0.0055(-0.0113)}^{+0.0057(+0.0112)}$ \\ \hline
    \end{tabular}
    \caption{Mean values and uncertainties at 1-$\sigma$ C.L. (2-$\sigma$ C.L. in parenthesis) for the cosmological parameters $H_0, \Omega_m$ and the derived one $\Omega_\alpha$. The $\Omega_\alpha$ represents the minimal coupling of the logarithmic function to the STEGR.}
    \label{tab:mean_values}
\end{table}

\begin{figure}
    \centering
    \includegraphics[scale=0.4]{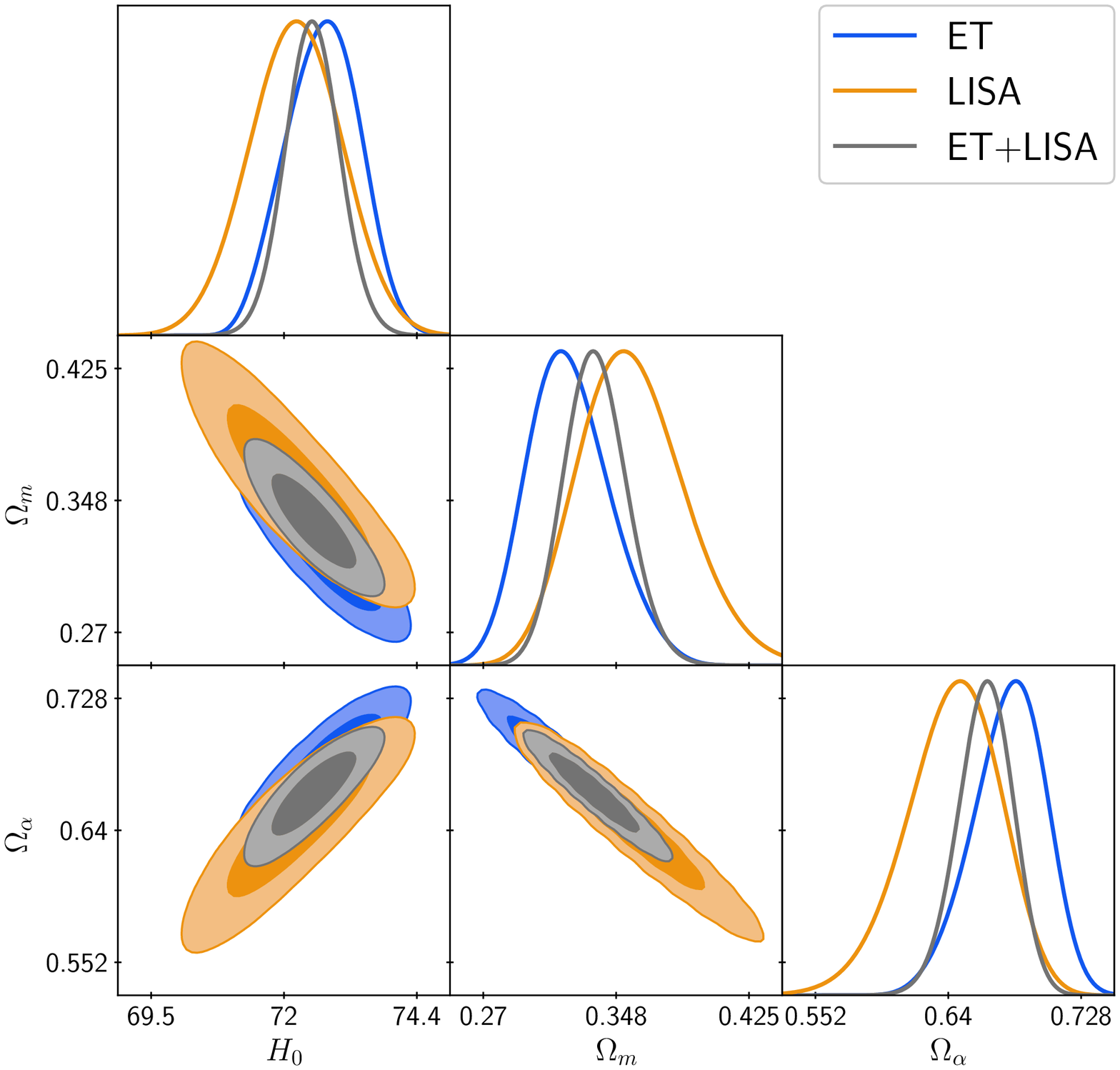}
       \includegraphics[scale=0.4]{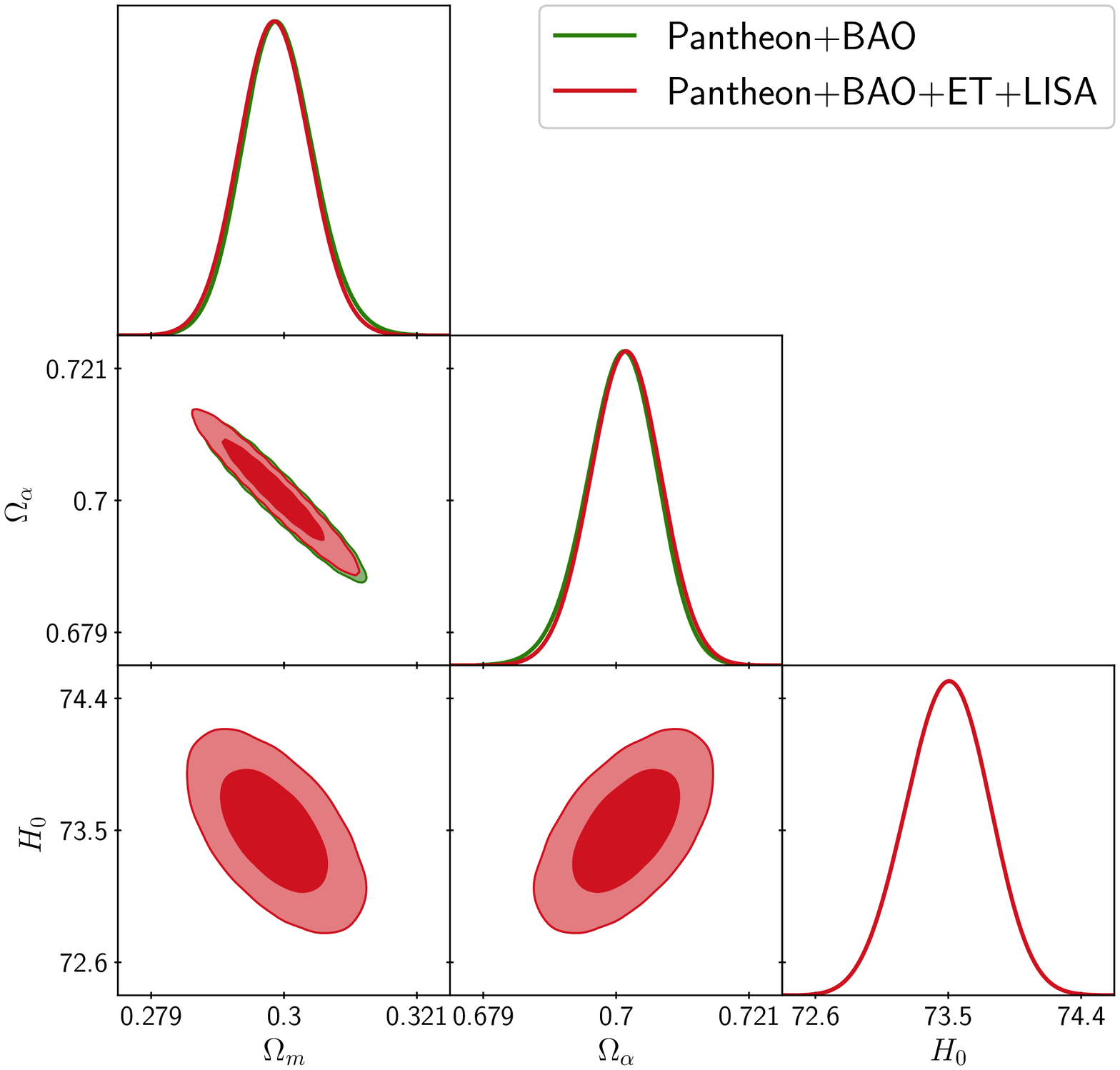}
    \caption{\textit{Left:} 1-$\sigma$ and 2-$\sigma$ confidence contours for the cosmological parameters $H_0, \Omega_M$ and the derived parameter $\Omega_\alpha$. We included the results for the following mock catalogs: ET, LISA, and the joint ET+LISA.
    \textit{Right:} 1-$\sigma$ and 2-$\sigma$ confidence contours for the cosmological parameters $H_0, \Omega_M$ and the derived parameter $\Omega_\alpha$. We included the results for the following catalogs: the joint Pantheon+BAO and the joint Pantheon+BAO+ET+LISA including GW mock data.
    }
    \label{fig:simulated_triangle}
\end{figure}

We present the results for the 1-$\sigma$ and 2-$\sigma$ confidence contours in Figure \ref{fig:simulated_triangle}. We report one plot including only simulated baselines (ET, LISA, and the joint ET+LISA) and one plot including the observational baseline and its comparison with the complete observational plus simulated baselines. \\

As we can see in Table \ref{tab:mean_values}, the Pantheon+BAO baseline can constraint $\Omega_m$ with a relative error of 2.03\%. Starting from this value of $\Omega_m = 0.2993$, and the $H_0$ value of the SH0ES collaboration \cite{riess2022comprehensive}, we generated the simulated baselines. Therefore, the constraints are expected to be consistent with the parameter vector $(H_0, \Omega_m ) = (73.04, 0.2993)$. We can notice in Table \ref{tab:mean_values} and Figure \ref{fig:simulated_triangle} that MCMC values are consistent at 1-$\sigma$ C.L. for ET and at 2-$\sigma$ for LISA and ET+LISA. Hence, consistency is achieved. It is interesting to see that the $H_0$ constraint for the full baseline Pantheon+BAO+ET+LISA is noticeably higher than for the other cases. Even so, it is consistent with the calibrated $H_0$ at 2-$\sigma$ C.L. \\

We can explore the relative error with the simulated baselines. For the ET, the relative error for $H_0$ is 1.01\% while for $\Omega_m$ is 8.63\%. This relative error is similar to the one found in \cite{escamilla2022dynamical}, where it was mentioned that we can expect to need around 1000 standard sirens to constraint $H_0$ with a relative error of around 1\%. For LISA, the relative error for $H_0$ is 1.21\% while it is 9.7\% for $\Omega_m$. This is a similar performance to ET with a much smaller baseline. This can be explained by the fact that the uncertainties of the simulated $d_L^{\text{gw}}$ are substantially smaller for LISA than for ET as we can see from Figure \ref{fig:logarithmicRatio}. The cause of this phenomenon comes from the bigger masses of the MBHB causing the standard sirens detected by LISA. Finally, for the joint baseline ET+LISA, the relative error in $H_0$ is 0.70\% and 5.78\% for $\Omega_m$. This represents an amazing precision with which we will be able to test modified gravity models. We will do so once the ET and LISA are in operation and with enough standard sirens as the ones simulated here. 

\subsection{About the $f(Q)$ logarithmic model}

It is now interesting to look at the derived parameter $\Omega_\alpha$. Since we are considering a $f(Q)$ gravity model given by Eq.(\ref{eqn:f(q)generalmodel}), the derived parameter $\Omega_\alpha$, quantifies the strength of the deviation from the STEGR. The higher the $\alpha$ (or $\Omega_\alpha$), the higher the deviation from GR. Furthermore, this parameter acts as a geometrical source of accelerated expansion as we described in Sec~\ref{sec:fQmodel}. Therefore, we expect that it plays a similar role to $\Lambda$ (however, $\Omega_\alpha \neq \Omega_\Lambda$ since lambda is the density parameter of an exotic fluid while $\Omega_\alpha$ comes from the minimal coupling of the STEGR to a logarithmic term in the non-metricity scalar $Q$). To make this comparison, we will focus on the constraint value for Pantheon+BAO, since it is the only one coming from truly observational baselines. As we can see in Table \ref{tab:mean_values}, we have $\Omega_\alpha = 0.7005^{+0.0061}_{-0.0052}$, which is very similar to the value determined from the $\Lambda$CDM model and SNIa only of $\Omega_\Lambda = 0.702 \, \pm \, 0.022$ \cite{Pan-STARRS1:2017jku}. This allows us to see that with this observational baseline, a universe governed by the logarithmic $f(Q)$ model has a similar matter content in comparison with $\Lambda$CDM. However, for this MG model, the cause of the cosmic accelerated expansion comes from the minimal coupling of this geometrical logarithmic term instead of an exotic fluid with negative pressure. This confirms that a universe with a logarithmic symmetric teleparallel coupling can explain this cosmic acceleration after being tested with actual observational data. We can also compute the value of the deceleration parameter with the observational baseline, which gives $q_0 = -0.6675^{+0.0064}_{-0.0075}$. For $\Lambda$CDM and SNIa only, is $q_0 = -0.553\,\pm\,0.33$. Notice that the coupling of the logarithmic term to the STEGR brings a higher accelerated expansion at late times. \\

On the other hand, with the aid of the simulated GW baselines, we can explore the predicted deviation from GR and hence, how the future ET and LISA standard sirens will help us to determine whether this $f(Q)$ model is feasible. In Figure \ref{fig:logarithmicRatio}, we present the evolution of the ratio $d_L^{\text{gw}}(z)/d_L^{\text{em}}(z)$ with redshift $z$. Notice that this ratio decreases rapidly with $z$. Since the difference between $d_L^{\text{gw}}$ and $d_L^{\text{em}}$ quantifies the deviation from GR, this $f(Q)$ model has a significant deviation from GR. 
This deviation is higher than 13\% at $z=1$, and it continues growing to reach a deviation of more than 18\% in the median value. 

\section{Conclusions}
\label{sec:conclusions}

In this work, we studied an interesting $f(Q)$ gravity model in the framework of symmetric teleparallel gravity. This model successfully predicted the cosmic late-time accelerated expansion in geometrical means when testing it with SNeIa and BAO data. The model is given by $f(Q) = Q/(8\pi G) - \alpha \ln(Q/Q_0)$ and it includes the STEGR minimally coupled to a logarithmic term that causes the accelerated expansion at late times. From the observational baseline (Pantheon+BAO), the derived deceleration parameter is given by $q_0 = -0.6675^{+0.0064}_{-0.0075}$ and the matter content $\Omega_m = 0.2993^{+0.0052}_{-0.0061}$. For the derived parameter, $\Omega_\alpha$, which quantifies the deviation from GR and causes the geometrical accelerated expansion, we get $\Omega_\alpha = 0.7005^{+0.0061}_{-0.0052}$, which is similar to the result from Pantheon and $\Lambda$CDM of $\Omega_\Lambda = 0.702 \, \pm \, 0.022$ \cite{Pan-STARRS1:2017jku}. This shows that $\Omega_\alpha$ is playing the role of the cosmological constant in the background perspective but it is not dark energy but a parameter coming from non-metricity geometry. Furthermore, the WEC, NEC, and DEC energy conditions were fulfilled while the SEC was violated under such values, which corroborates that this model is capable of predicting a universe with accelerating expansion. This result is a good input in order to consider this model as a good candidate to solve the cosmological constant problem.\\

Furthermore, we generated two simulated standard siren baselines assuming the cosmology given by the logarithmic $f(Q)$ model. This allowed us to study the differences between the logarithmic $f(Q)$ model and GR in a tensor perturbation approach. In particular, to generate our baselines, we considered the performance of the planned ET and LISA detectors. With the use of 1096 simulated samples, we got the constraint of the Hubble constant $H_0$ and the matter content $\Omega_m$ with relative errors of 0.71\% and 5.78\%, respectively. This shows that once these detectors are in operation, we are going to be capable of accurately testing modified gravity theories. Moreover, with the aid of standard sirens, it will be possible to quantify deviations from GR. We computed the percentile deviation expected from the $f(Q)$ logarithmic gravity model. The ratio $d^{gw}_L(z)/d^{em}_L(z)$is expected to have deviations of more than 13\% at $z=1$ and more than 18\% for high redshift, where ET and LISA are going to be capable of detecting standard sirens. \\

The significant deviation found from our analysis was not reported in \cite{d2022forecasting} when studying a $f(Q)$ power-law model. For this power-law model, the evolution is consistent with GR at 1-$\sigma$. However, the only current standard siren, GW170817 \cite{abbott2017gw170817} having a confirmed electromagnetic counterpart GRB-170817A, has a tiny redshift of $z=0.0099$ \cite{abbott2017gravitational}. Then, from current data, it is not possible to determine whether a huge GW deviation is physically possible. Thus, only future GW data will find out if a model like the one presented in this work is realistic and feasible. \\

\acknowledgments
JAN acknowledges financial support from the ``Excellence Project Scholarship'' funded by the University of Padova. 
CE-R acknowledges the Royal Astronomical Society as FRAS 10147 and is supported by PAPIIT UNAM Project TA100122. 
This article/publication is based upon work from COST Action CA21136 Addressing observational tensions in cosmology with systematics and fundamental physics (CosmoVerse) supported by COST (European Cooperation in Science and Technology). 

\bibliographystyle{JHEP}
\bibliography{references}
\end{document}